\documentclass{article}
\usepackage{amssymb}

\input{tcilatex}

\begin{document}

\title{Continuum variable entangled state generated by an asymmetric beam
splitter }
\author{$^{1,2}$Hong-yi Fan and $^{1}$Yan-li Yang \\
%EndAName
$^{1}$Department of Material Science and Engineering, University of\\
Science\\
and Technology of China, Hefei, Anhui 230026, China\\
$^{2}$Department of Physics, Shanghai Jiao Tong University, Shanghai 200030,%
\\
China}
\maketitle

\begin{abstract}
The generating entangled state by using a 50/50 beamsplitter has been
discussed in the literature before . In this paper we explore how to use an
asymmetric beam-splitter to produce a new kind of entangled state. We
construct such kind of states theoretically and then prove that they make up
a complete and orthonormal representation in two-mode Fock space. Its
application in finding new squeezing operator and new squeezed state is
introduced.
\end{abstract}

\section{Introduction}

Recently, quantum entanglement, which originated from Einstein, Podolsky and
Rosen (EPR) in a paper arguing the incompleteness of quantum mechanics,$^{1}$
is of increasingly interest in studies of quantum information and quantum
communication. It lies at the core of some new applications in the emerging
field of quantum communication science $^{2-7}$. The concept of entanglement
has played a key role in understanding some fundamental problems in quantum
mechanics and quantum optics. In an quantum entangled state, a measurement
performed on one part of the system provides information on the remaining
part, this has now been known as a basic feature of quantum mechanics,
though it seems weird. Thus an entangled composite system is nonseparable.
In EPR's pioneer argument, the entanglement was revealed by explicitly
writing the wave function of a bipartite with their relative position $%
X_{1}-X_{2\text{ }}$being $x_{0}$ and their total momentum $P_{1}+P_{2}\;$%
being$\;p_{0}=0,$ i.e. $\psi (x_{1},x_{2})=\frac{1}{2\pi }\int_{-\infty
}^{\infty }dpe^{ip\left( x_{1}-x_{2}+x_{0}\right) }.$ In Ref. [8] the
simultaneous eigenstate $\left\vert \eta \right\rangle $ of commutative
operators $\left( X_{1}-X_{2},P_{1}+P_{2}\right) $ in the two-mode Fock
space is found, 
\begin{equation}
\left\vert \eta \right\rangle =\exp [-\frac{1}{2}\left\vert \eta \right\vert
^{2}+\eta a_{1}^{\dagger }-\eta ^{\ast }a_{2}^{\dagger }+a_{2}^{\dagger
}a_{1}^{\dagger }]\left\vert 00\right\rangle _{12},  \label{1}
\end{equation}%
where $\eta =(\eta _{1}+i\eta _{2})/\sqrt{2}$\ is a complex number, $%
\left\vert 00\right\rangle \;$is the two-mode vacuum state, ($%
a_{i},a_{i}^{\dagger }),$ $i=1,2,\;$are two-mode Bose annihilation and
creation operators in Fock space, related to $\left( X_{i},P_{i}\right) \;$%
by $X_{i}=(a_{i}+a_{i}^{\dagger })/\sqrt{2},\;P_{i}=(a_{i}-a_{i}^{\dagger
})/\left( \sqrt{2}i\right) .$ The basic ingredient of the $\left\vert \eta
\right\rangle $ state about the coordinate-momentum entanglement can be
demonstrated through its disentangling process, 
\begin{equation}
\left\vert \eta =(\eta _{1}+i\eta _{2})/\sqrt{2}\right\rangle =e^{-i\eta
_{1}\eta _{2}/2}\int\limits_{-\infty }^{\infty }dx\left\vert x\right\rangle
_{1}\otimes \left\vert x-\eta _{1}\right\rangle _{2}e^{ix\eta _{2}},
\label{2}
\end{equation}%
where $\left\vert x\right\rangle _{i}\;$is the coordinate eigenstate of $%
X_{i},$%
\begin{equation}
\left\vert x\right\rangle _{i}=\pi ^{-1/4}\exp [-\frac{1}{2}x^{2}+\sqrt{2}%
xa_{i}^{\dagger }-\frac{1}{2}a_{i}^{\dagger 2}]\left\vert 0\right\rangle
_{i}.\;\;  \label{3}
\end{equation}%
Eq. (2) shows that once particle 1 is measured in the state $\left\vert
x\right\rangle _{1},$ particle 2 immediately collapses to the coordinate
eigenstate $\left\vert x-\eta _{1}\right\rangle _{2}.$ Eq. (2) is named
Schmidt decomposition according to Ref. [9]. On the other hand, the Schmidt
decomposition of $\left\vert \eta \right\rangle $ in the two-mode momentum
basis is 
\begin{equation}
\left\vert \eta \right\rangle =e^{i\eta _{1}\eta _{2}/2}\int_{-\infty
}^{\infty }dp\left\vert p\right\rangle _{1}\otimes \left\vert \eta
_{2}-p\right\rangle _{2}e^{-i\eta _{1}p},  \label{4}
\end{equation}%
where $\left\vert p\right\rangle _{i}$ is the momentum eigenvector of $%
P_{i}, $ 
\begin{equation}
\left\vert p\right\rangle _{i}=\pi ^{-1/4}\exp [-\frac{1}{2}p^{2}+i\sqrt{2}%
pa_{i}^{\dagger }+\frac{1}{2}a_{i}^{\dagger 2}]\left\vert 0\right\rangle
_{i},  \label{5}
\end{equation}%
which tells us that once particle 1 is measured in the state $\left\vert
p\right\rangle _{1}$, particle 2 immediately collapses to the momentum
eigenstate $\left\vert \eta _{2}-p\right\rangle _{2}$ no matter how far the
distance between the two particles is. Thus (2) and (4) together implies the
quantum entanglement. Note that the $\left\vert \eta \right\rangle $ states
obey the eigenvector equations 
\begin{equation}
\left( a_{1}-a_{2}^{\dagger }\right) \left\vert \eta \right\rangle =\eta
\left\vert \eta \right\rangle ,\;\;\ \left( a_{2}-a_{1}^{\dagger }\right)
\left\vert \eta \right\rangle =-\eta ^{\ast }\left\vert \eta \right\rangle .
\label{6}
\end{equation}%
It then follows 
\begin{equation}
\left( X_{1}-X_{2}\right) \left\vert \eta \right\rangle =\eta _{1}\left\vert
\eta \right\rangle ,  \label{7}
\end{equation}%
\begin{equation}
\left( P_{1}+P_{2}\right) \left\vert \eta \right\rangle =\eta _{2}\left\vert
\eta \right\rangle .  \label{8}
\end{equation}%
The experimental implementation of entangled state of continuous variables
does not use the position and momentum of particles but uses light beams
that can be characterized by parameters obeying the same commutation
relations as position operator $X_{i}$ and momentum operator $P_{i}$. The
analogy is based on the fact that a single mode of the quantized radiation
field can be expressed in terms of annihilation operators $a_{i}$ and
creation operator $a_{i}^{\dagger }$ of a quantum harmonic oscillator with
frequency $\omega $, i.e. the electric field operator can be described as $%
E_{i}\sim X_{i}\cos \omega t+P_{i}\sin \omega t$ . It is now known that the
EPR light fields with bipartite entanglement can be built from
two-single-mode squeezed vacuum state combined at a 50/50 beam splitter $%
^{10}$, i.e. two light fields maximally squeezed in $X_{i}\ $and $P_{i}$ (in
opposite quadratures)$,$ respectively entering the two input ports of a
50/50 beamsplitter produce at the output of the beamsplitter a pair of
entangled light beams. It is also known that even one single-mode squeezed
state incident on a beam splitter yields a bipartite entangled state,
because the quantized vacuum field also enters in another input port of the
beam splitter and contributes to the two output modes $^{11}$.

An interesting and practical question thus naturally arises: if the
beamsplitter is not a 50/50 one, but an asymmetric one, then what is the
output state when two light fields maximally squeezed in $X_{i}\ $and $%
P_{i}, $ respectively entering its two input ports and get superimposed? For
an asymmetric beamsplitter without absorption within itself, its complex
amplitude reflectivity $r$ and transmissivity $t$ for light incident from
one side (or $r^{\prime },$ $t^{\prime }$ for light coming from the other
side) are not equal to each other. The incident fields ($a_{1}$ and $a_{2}$%
), the reflected field $a_{3}$ and the transmitted field $a_{4}$ may be
related by a \textquotedblleft scattering matrix\textquotedblright\ $^{11}$ 
\begin{equation}
\left( 
\begin{array}{c}
a_{3} \\ 
a_{4}%
\end{array}%
\right) =\left( 
\begin{array}{cc}
t^{\prime } & r \\ 
r^{\prime } & t%
\end{array}%
\right) \left( 
\begin{array}{c}
a_{1} \\ 
a_{2}%
\end{array}%
\right) ,  \label{9}
\end{equation}%
where $t,\;r,\;t^{\prime },$ and $r^{\prime }$ obey the reciprocity
relations 
\begin{equation}
\left\vert r^{\prime }\right\vert =\left\vert r\right\vert ,\;\left\vert
t^{\prime }\right\vert =\left\vert t\right\vert ,\;\text{\ }\left\vert
r\right\vert ^{2}+\left\vert t\right\vert ^{2}=1,\text{ }\;r^{\ast
}t^{\prime }+r^{\prime }t^{\ast }=0,\text{ }\;r^{\ast }t+r^{\prime
}t^{\prime \ast }=0,  \label{10}
\end{equation}%
or the role of a beam splitter operation on two input modes is equivalent to
the unitary operator $B\equiv \exp \left[ \theta (a_{1}^{\dagger
}a_{2}-a_{2}^{\dagger }a_{1})\right] ,\;\theta \neq 0,$ (we do not consider
the phase difference between the reflected and transmitted fields), with the
amplitude reflection and transmission coefficients $t=\cos \theta ,\;r=\sin
\theta .$ The role of $B$ is $Ba_{1}B^{-1}=a_{3},$ $Ba_{2}B^{-1}=a_{4}.$ The
details of relationship between two input modes and two output modes for the
beam splitter is discussed in [11]. In this work we want to derive the
output state for the asymmetric beamsplitter, which turns out to be a new
entangled state characteristic of $\theta .$ Then we study its main
properties and present its application. Our work is arranged as follows: In
Sec. 2 and 3 we construct the new two-mode entangled state, denoted as $%
\left\vert \eta ,\theta \right\rangle ,$ which experimentally can be
generated by an asymmetric beamsplitter, i.e. two light fields maximally
squeezed in opposite quadratures$,$ respectively entering two input ports of
a non-50/50 beamsplitter and get superimposed, will produce at the output a
pair of entangled light beams expressed by $\left\vert \eta ,\theta
\right\rangle .$ In Sec. 4 we discuss the orthonomal and completeness
relation of $\left\vert \eta ,\theta \right\rangle $ and calculate the
weight factor for the completeness. In Sec. 5 and 6 we show how to apply $%
\left\vert \eta ,\theta \right\rangle $ to deriving some new generalized
squeezed states.

\section{The new entangled state $\left\vert \protect\eta ,\protect\theta %
\right\rangle $}

In the case when two light fields maximally squeezed in $X_{i}\ $and $P_{i},$
respectively entering a beam-splitter's two input ports and get
superimposed, we find that the output state emerging from asymmetric
beam-splitter is 
\begin{equation}
\begin{array}{c}
\left\vert \eta ,\theta \right\rangle =\exp \{-\frac{1}{2}\left\vert \eta
\right\vert ^{2}+\eta a_{1}^{\dagger }-\eta ^{\ast }(a_{2}^{\dagger }\sin
2\theta +a_{1}^{\dagger }\cos 2\theta ) \\ 
+\frac{1}{2}\eta ^{\ast 2}\cos 2\theta +a_{1}^{\dagger }a_{2}^{\dagger }\sin
2\theta +\frac{1}{2}(a_{1}^{\dagger 2}-a_{2}^{\dagger 2})\cos 2\theta
\}\left\vert 00\right\rangle .%
\end{array}
\label{11}
\end{equation}%
Clearly, when $\theta =\pi /4,$ which corresponds to a 50/50 beam-splitter, $%
\left\vert \eta ,\pi /4\right\rangle $ reduces to $\left\vert \eta
\right\rangle $. However, it must be clarified that $\left\vert \eta ,\theta
\right\rangle $ is not a rotated state of $\left\vert \eta \right\rangle $,
i.e., 
\begin{equation}
\left\vert \eta ,\theta \right\rangle \neq \exp \left[ \theta
(a_{1}^{\dagger }a_{2}\pm a_{2}^{\dagger }a_{1})\right] \left\vert \eta
\right\rangle .  \label{12}
\end{equation}%
Operating $a_{i},$ $i=1,2,$ on $\left\vert \eta ,\theta \right\rangle $
respectively gives 
\begin{equation}
\left( a_{1}-a_{2}^{\dagger }\sin 2\theta -a_{1}^{\dagger }\cos 2\theta
\right) \left\vert \eta ,\theta \right\rangle =\left( \eta -\eta ^{\ast
}\cos 2\theta \right) \left\vert \eta ,\theta \right\rangle ,  \label{13}
\end{equation}%
and 
\begin{equation}
\left( a_{2}-a_{1}^{\dagger }\sin 2\theta +a_{2}^{\dagger }\cos 2\theta
\right) \left\vert \eta ,\theta \right\rangle =-\eta ^{\ast }\sin 2\theta
\left\vert \eta ,\theta \right\rangle .  \label{14}
\end{equation}%
From (13)-(14) we can deduce 
\begin{equation}
\left( a_{1}\sin 2\theta -a_{2}\cos 2\theta -a_{2}^{\dagger }\right)
\left\vert \eta ,\theta \right\rangle =\eta \sin 2\theta \left\vert \eta
,\theta \right\rangle ,  \label{15}
\end{equation}%
and 
\begin{equation}
\left( a_{1}\cos 2\theta +a_{2}\sin 2\theta -a_{1}^{\dagger }\right)
\left\vert \eta ,\theta \right\rangle =\left( \eta \cos 2\theta -\eta ^{\ast
}\right) \left\vert \eta ,\theta \right\rangle .  \label{16}
\end{equation}%
Subtracting (16) from (13) yields 
\begin{equation}
\left( X_{2}-X_{1}\tan \theta \right) \left\vert \eta ,\theta \right\rangle
=-\eta _{1}\tan \theta \left\vert \eta ,\theta \right\rangle ,  \label{17}
\end{equation}%
adding (14) and (15) leads to%
\begin{equation}
\left( P_{1}+P_{2}\tan \theta \right) \left\vert \eta ,\theta \right\rangle
=\eta _{2}\left\vert \eta ,\theta \right\rangle ,  \label{18}
\end{equation}%
so $\left\vert \eta ,\theta \right\rangle $ is the common eigenvector of $%
\left( X_{2}-X_{1}\tan \theta \right) $ and $\left( P_{1}+P_{2}\tan \theta
\right) .$ When $\theta =\frac{\pi }{4},$ (17)-(18) reduce to (7)-(8).
Therefore, $\left\vert \eta ,\theta \right\rangle $ is a new enangled state
with a non-trivial expression (see (11)) and one can Schmidt-decompose it
too.

\section{The physical meaning of $\left\vert \protect\eta ,\protect\theta %
\right\rangle \ $and its relation to an asymmetric beamsplitter}

We now explain why the state $\left\vert \eta ,\theta \right\rangle $ can
describe the production of new entangled light fields using two maximally
squeezed light fields in opposite directions (respectively represented by $%
\left\vert p=0\right\rangle _{1}$ and $\left\vert x=0\right\rangle _{2})$
and a non-50/50 beamsplitter, Let the assymetric beam splitter operator be $%
\exp \left[ 2\theta (a_{2}^{\dagger }a_{1}-a_{1}^{\dagger }a_{2})\right]
\equiv \exp [-2i\theta J_{y}]$, from 
\begin{equation}
\begin{array}{c}
\exp \left[ -2i\theta J_{y}\right] a_{1}^{\dagger }\exp \left[ 2i\theta J_{y}%
\right] =a_{1}^{\dagger }\cos \theta +a_{2}^{\dagger }\sin \theta , \\ 
\exp \left[ -2i\theta J_{y}\right] a_{2}^{\dagger }\exp \left[ 2i\theta J_{y}%
\right] =a_{2}^{\dagger }\cos \theta -a_{1}^{\dagger }\sin \theta ,%
\end{array}
\label{19}
\end{equation}%
and (3) and (5) we have 
\begin{eqnarray}
&&\exp \left[ 2\theta (a_{2}^{\dagger }a_{1}-a_{1}^{\dagger }a_{2})\right]
\left\vert p=0\right\rangle _{1}\otimes \left\vert x=0\right\rangle _{2}
\label{20} \\
&=&\exp \left[ a_{1}^{\dagger }a_{2}^{\dagger }\sin 2\theta +\frac{1}{2}%
\left( a_{1}^{\dagger 2}-a_{2}^{\dagger 2}\right) \cos 2\theta \right]
\left\vert 00\right\rangle =\left\vert \eta =0,\theta \right\rangle . 
\nonumber
\end{eqnarray}%
Then operating the displacement operator $D_{1}(\eta )\equiv \exp [\eta
a_{1}^{\dagger }-\eta ^{\ast }a_{1}]$ on (20) leads to (11), i.e. 
\begin{equation}
\begin{array}{c}
D_{1}(\eta )\exp \left[ a_{2}^{\dagger }a_{1}^{\dagger }\sin 2\theta +\frac{1%
}{2}\left( a_{1}^{\dagger 2}-a_{2}^{\dagger 2}\right) \cos 2\theta \right]
\left\vert 00\right\rangle \\ 
=\exp \{-\frac{1}{2}\left\vert \eta \right\vert ^{2}+\eta a_{1}^{\dagger
}-\eta ^{\ast }(a_{2}^{\dagger }\sin 2\theta +a_{1}^{\dagger }\cos 2\theta )+%
\frac{1}{2}\eta ^{\ast 2}\cos 2\theta \\ 
+a_{1}^{\dagger }a_{2}^{\dagger }\sin 2\theta +\frac{1}{2}\left(
a_{1}^{\dagger 2}-a_{2}^{\dagger 2}\right) \cos 2\theta \}\left\vert
00\right\rangle =\left\vert \eta ,\theta \right\rangle .%
\end{array}
\label{21}
\end{equation}%
Experimentally, this displacement can be implemented by reflecting the light
field of $\left\vert \eta =0,\theta \right\rangle $ from a partially
reflecting mirror (say 99\% reflection and 1\% transmission) and adding
through the mirror a field that has been phase and amplitude modulated
according to the value $\eta \equiv |\eta |e^{i\Phi }$.

\section{The properties of $\left| \protect\eta ,\protect\theta %
\right\rangle $}

We now examine the main properties of $\left\vert \eta ,\theta \right\rangle
.$ Using the mathematical formula 
\begin{eqnarray}
\int \frac{d^{2}z}{\pi }\exp \{\zeta \left\vert z\right\vert ^{2}+\xi z+\eta
z^{\ast }+fz^{2}+gz^{\ast 2}\} &=&\frac{1}{\sqrt{\zeta ^{2}-4fg}}\exp \left[ 
\frac{-\zeta \xi \eta +\xi ^{2}g+\eta ^{2}f}{\zeta ^{2}-4fg}\right] ,
\label{22} \\
Re(\zeta +f+g) &<&0,\;\;Re\left( \frac{\zeta ^{2}-4fg}{\zeta +f+g}\right)
<0,\;  \nonumber \\
\text{or \ }Re(\zeta -f-g) &<&0,\;\;Re\left( \frac{\zeta ^{2}-4fg}{\zeta -f-g%
}\right) <0,  \nonumber
\end{eqnarray}%
where $\zeta ,$ $f,$ $g$ are so selected as to insure the integration
convergent, and using the normal ordered form of the vacuum projector (: :
denotes normal ordering), 
\begin{equation}
\left\vert 00\right\rangle \left\langle 00\right\vert =:\exp
\{-a_{1}^{\dagger }a_{1}-a_{2}^{\dagger }a_{2}\}:,  \label{23}
\end{equation}%
as well as the technique of integration within an ordered product (IWOP) of
operators $^{12-13}$ we can prove that $\left\vert \eta ,\theta
\right\rangle $ expressed by (11) make up a complete set, i.e., 
\begin{equation}
\begin{array}{c}
\sin 2\theta \dint \frac{d^{2}\eta }{\pi }\left\vert \eta ,\theta
\right\rangle \left\langle \eta ,\theta \right\vert \\ 
=\sin 2\theta \dint \frac{d^{2}\eta }{\pi }:\exp \{-\left\vert \eta
\right\vert ^{2}+\eta \left( a_{1}^{\dagger }-a_{2}\sin 2\theta -a_{1}\cos
2\theta \right) \\ 
+\eta ^{\ast }\left( a_{1}-a_{2}^{\dagger }\sin 2\theta -a_{1}^{\dagger
}\cos 2\theta \right) +\frac{1}{2}\left( \eta ^{2}+\eta ^{\ast 2}\right)
\cos 2\theta \\ 
+\left( a_{1}^{\dagger }a_{2}^{\dagger }+a_{1}a_{2}\right) \sin 2\theta +%
\frac{1}{2}\left( a_{1}^{\dagger 2}-a_{2}^{\dagger
2}+a_{1}^{2}-a_{2}^{2}\right) \cos 2\theta -a_{1}^{\dagger
}a_{1}-a_{2}^{\dagger }a_{2}\}: \\ 
=:\exp \{\frac{1}{\sin ^{2}2\theta }[\left( a_{1}^{\dagger }-a_{2}\sin
2\theta -a_{1}\cos 2\theta \right) \left( a_{1}-a_{2}^{\dagger }\sin 2\theta
-a_{1}^{\dagger }\cos 2\theta \right) \\ 
+\frac{1}{2}\cos 2\theta \left( a_{1}^{\dagger }-a_{2}\sin 2\theta
-a_{1}\cos 2\theta \right) ^{2}+\frac{1}{2}\cos 2\theta \left(
a_{1}-a_{2}^{\dagger }\sin 2\theta -a_{1}^{\dagger }\cos 2\theta \right)
^{2}] \\ 
+\left( a_{1}^{\dagger }a_{2}^{\dagger }+a_{1}a_{2}\right) \sin 2\theta +%
\frac{1}{2}\left( a_{1}^{\dagger 2}-a_{2}^{\dagger
2}+a_{1}^{2}-a_{2}^{2}\right) \cos 2\theta -a_{1}^{\dagger
}a_{1}-a_{2}^{\dagger }a_{2}\}: \\ 
=:e^{0}:=1.%
\end{array}
\label{24}
\end{equation}%
Here the factor $\sin 2\theta $ is needed for the completeness relation. One
might think the fact that the two-mode states at the output of the
beam-splitter form a complete basis is trivial, now form the derivation of
(24) we see the integration weight factor $\sin 2\theta $ is not a trivial
one, it is a result of really performing the integration $\dint \frac{%
d^{2}\eta }{\pi }\left\vert \eta ,\theta \right\rangle \left\langle \eta
,\theta \right\vert .$ From (15) and the Hermite conjugate of (14) we have 
\begin{equation}
\left\langle \eta ^{\prime },\theta \right\vert \left( a_{1}\sin 2\theta
-a_{2}\cos 2\theta -a_{2}^{\dagger }\right) \left\vert \eta ,\theta
\right\rangle =\eta \sin 2\theta \left\langle \eta ^{\prime },\theta \right.
\left\vert \eta ,\theta \right\rangle =\eta ^{\prime }\sin 2\theta
\left\langle \eta ^{\prime },\theta \right. \left\vert \eta ,\theta
\right\rangle .  \label{25}
\end{equation}%
It then follows%
\begin{equation}
\sin 2\theta \left( \eta -\eta ^{\prime }\right) \left\langle \eta ^{\prime
},\theta \right. \left\vert \eta ,\theta \right\rangle =0.  \label{26}
\end{equation}%
Similarly, from (16) and the Hermite conjugate of (13) we derive 
\begin{equation}
\begin{array}{c}
\left\langle \eta ^{\prime },\theta \right\vert \left( a_{1}\cos 2\theta
+a_{2}\sin 2\theta -a_{1}^{\dagger }\right) \left\vert \eta ,\theta
\right\rangle =\left( \eta \cos 2\theta -\eta ^{\ast }\right) \left\langle
\eta ^{\prime },\theta \right\vert \left. \eta ,\theta \right\rangle \\ 
=\left( \eta ^{\prime }\cos 2\theta -\eta ^{\prime \ast }\right)
\left\langle \eta ^{\prime },\theta \right\vert \left. \eta ,\theta
\right\rangle , \\ 
\left[ \cos 2\theta \left( \eta -\eta ^{\prime }\right) +\left( \eta
^{\prime \ast }-\eta ^{\ast }\right) \right] \left\langle \eta ^{\prime
},\theta \right. \left\vert \eta ,\theta \right\rangle =0.%
\end{array}
\label{27}
\end{equation}%
Combining the results of (25)-(26) we obtain 
\begin{equation}
\tan 2\theta \left( \eta ^{\prime \ast }-\eta ^{\ast }\right) \left\langle
\eta ^{\prime },\theta \right. \left\vert \eta ,\theta \right\rangle =0.
\label{28}
\end{equation}%
As a consequence of (25) and (28) and in reference to (24) we conclude 
\begin{equation}
\left\langle \eta ^{\prime },\theta \right. \left\vert \eta ,\theta
\right\rangle =2\pi \delta \left( \eta _{1}-\eta _{1}^{\prime }\right)
\delta \left( \eta _{2}-\eta _{2}^{\prime }\right) /\sin 2\theta ,\text{\ \ }%
\eta =(\eta _{1}+i\eta _{2})/\sqrt{2}.  \label{29}
\end{equation}%
According to Dirac's theory on representation in qauntum mechanics, the set
of $\left\vert \eta ,\theta \right\rangle $ make up a new orthonormal and
complete representation in the two-mode Fock space, which is an another
entangled state representation. For a review of various applications of the
EPR entangled state representation of continuum variables we refer to [14].

\section{The squeezing of $\left\vert \protect\eta ,\protect\theta %
\right\rangle $ and the corresponding squeezing operator}

As an application of the $\left\vert \eta ,\theta \right\rangle $
representation, now we construct the following ket-bra operator in an
integration form 
\begin{equation}
U=\sin 2\theta \dint \frac{d^{2}\eta }{\mu \pi }\left\vert \eta /\mu ,\theta
\right\rangle \left\langle \eta ,\theta \right\vert .  \label{30}
\end{equation}%
where $\eta \rightarrow \eta /\mu $ is a c-number dilation transformation.
We shall point out that $U$ is a new 2-mode squeezing operator (for a review
of squeezed states we refer to [15]). Letting $\mu =e^{\lambda },$ and using
(23) as well as the IWOP technique to perform this integration, we find the
normal ordering of $U$ is 
\begin{equation}
\begin{array}{c}
U=\sin 2\theta \dint \frac{d^{2}\eta }{\mu \pi }:\exp \{-\frac{1}{2}%
\left\vert \eta \right\vert ^{2}\left( 1+\frac{1}{\mu ^{2}}\right) \\ 
+\eta \left( a_{1}^{\dagger }/\mu -a_{2}\sin 2\theta -a_{1}\cos 2\theta
\right) +\eta ^{\ast }\left( a_{1}-a_{2}^{\dagger }\sin 2\theta /\mu
-a_{1}^{\dagger }\cos 2\theta /\mu \right) \\ 
+\frac{1}{2}\left( \frac{1}{\mu ^{2}}\eta ^{\ast 2}+\eta ^{2}\right) \cos
2\theta +\left( a_{1}^{\dagger }a_{2}^{\dagger }+a_{1}a_{2}\right) \sin
2\theta \\ 
+\frac{1}{2}\left( a_{1}^{\dagger 2}-a_{2}^{\dagger
2}+a_{1}^{2}-a_{2}^{2}\right) \cos 2\theta -a_{1}^{\dagger
}a_{1}-a_{2}^{\dagger }a_{2}\}: \\ 
=\frac{\sin 2\theta }{\sqrt{S}}\exp \{\frac{1}{2S}\sinh ^{2}\lambda \cos
2\theta (a_{1}^{\dagger 2}-a_{2}^{\dagger 2})+\frac{1}{2S}a_{1}^{\dagger
}a_{2}^{\dagger }\sinh 2\lambda \sin 2\theta \} \\ 
\times :\exp \{(a_{1}^{\dagger },a_{2}^{\dagger })\left( M-1\right) \left( 
\begin{array}{c}
a_{1} \\ 
a_{2}%
\end{array}%
\right) \}: \\ 
\times \exp \{\frac{1}{2S}\sinh ^{2}\lambda \cos 2\theta
(a_{1}^{2}-a_{2}^{2})-\frac{1}{2S}a_{1}a_{2}\sinh 2\lambda \sin 2\theta \},%
\end{array}
\label{31}
\end{equation}%
where we have set $S=\cosh ^{2}\lambda -\cos ^{2}2\theta ,$ and 
\begin{equation}
M=\frac{\sin 2\theta }{S}\left( 
\begin{array}{cc}
\cosh \lambda \sin 2\theta & \sinh \lambda \cos 2\theta \\ 
-\sinh \lambda \cos 2\theta & \cosh \lambda \sin 2\theta%
\end{array}%
\right) .  \label{32}
\end{equation}%
Especially, when $\theta =\pi /4,$%
\begin{equation}
\begin{array}{c}
U_{\theta =\pi /4}=\sec h\lambda \exp \{a_{1}^{\dagger }a_{2}^{\dagger
}\tanh \lambda \}:\exp \{(a_{1}^{\dagger }a_{1}+a_{2}^{\dagger }a_{2})\left( 
\text{sech}\lambda -1\right) :\exp \{-a_{1}a_{2}\tanh \lambda \} \\ 
=\dint \frac{d^{2}\eta }{\mu \pi }\left\vert \eta /\mu \right\rangle
\left\langle \eta \right\vert ,%
\end{array}
\label{33}
\end{equation}%
where $\left\vert \eta /\mu \right\rangle $ is given by (1), $U_{\theta =\pi
/4}$ is the usual two-mode squeezing operator. (33) indicates that the usual
two-mode squeezing operator has a neat representation in the entangled state
basis$^{16}$, this implies that two-mode squeezed state has close
relationship with the bipartite entangled state. No wonder the idler mode
and the signal mode, which come out of a parametric down-conversion
interaction and compose a two-mode squeezed state, are entangled in a
frequency domain. The matrix $M$ in (32) can be diagonalized as 
\begin{equation}
M=\frac{\sin 2\theta }{S}\left( 
\begin{array}{cc}
1/2 & i/2 \\ 
i/2 & 1/2%
\end{array}%
\right) \left( 
\begin{array}{cc}
\alpha & 0 \\ 
0 & \alpha ^{\ast }%
\end{array}%
\right) \left( 
\begin{array}{cc}
1 & -i \\ 
-i & 1%
\end{array}%
\right) ,  \label{34}
\end{equation}%
where 
\begin{equation}
\alpha =\cosh \lambda \sin 2\theta +i\sinh \lambda \cos 2\theta ,\;\;|\alpha
|=\sqrt{S},\;\;\alpha =\sqrt{S}e^{i\varphi },\text{ \ }\varphi =\tan
^{-1}\left( \tanh \lambda \cot 2\theta \right) ,  \label{35}
\end{equation}%
so 
\begin{equation}
\begin{array}{c}
\ln M=\ln \frac{\sin 2\theta }{S}+\left( 
\begin{array}{cc}
1/2 & i/2 \\ 
i/2 & 1/2%
\end{array}%
\right) \left( 
\begin{array}{cc}
\ln \sqrt{S}+i\varphi & 0 \\ 
0 & \ln \sqrt{S}-i\varphi%
\end{array}%
\right) \left( 
\begin{array}{cc}
1 & -i \\ 
-i & 1%
\end{array}%
\right) \\ 
=\left( 
\begin{array}{cc}
\ln \frac{\sin 2\theta }{\sqrt{S}} & \varphi \\ 
-\varphi & \ln \frac{\sin 2\theta }{\sqrt{S}}%
\end{array}%
\right) .%
\end{array}
\label{36}
\end{equation}%
Thus 
\begin{equation}
:\exp \{(a_{1}^{\dagger },a_{2}^{\dagger })\left( M-1\right) \left( 
\begin{array}{c}
a_{1} \\ 
a_{2}%
\end{array}%
\right) \}:=\exp \{(a_{1}^{\dagger },a_{2}^{\dagger })\left( 
\begin{array}{cc}
\ln \frac{\sin 2\theta }{\sqrt{S}} & \varphi \\ 
-\varphi & \ln \frac{\sin 2\theta }{\sqrt{S}}%
\end{array}%
\right) \left( 
\begin{array}{c}
a_{1} \\ 
a_{2}%
\end{array}%
\right) \}.  \label{37}
\end{equation}%
Using the operator identity 
\begin{equation}
\exp \left[ a_{i}^{\dagger }\Lambda _{ij}a_{j}\right] a_{l}\exp \left[
-a_{i}^{\dagger }\Lambda _{ij}a_{j}\right] =\left( e^{-\Lambda }\right)
_{lj}a_{j},  \label{38}
\end{equation}%
we have 
\begin{equation}
U\left( 
\begin{array}{c}
a_{1} \\ 
a_{2}%
\end{array}%
\right) U^{-1}=M^{-1}[\left( 
\begin{array}{c}
a_{1} \\ 
a_{2}%
\end{array}%
\right) -K\left( 
\begin{array}{c}
a_{1}^{\dagger } \\ 
a_{2}^{\dagger }%
\end{array}%
\right) ],  \label{39}
\end{equation}%
where 
\begin{equation}
M^{-1}=\left( 
\begin{array}{cc}
\cosh \lambda & -\cot 2\theta \sinh \lambda \\ 
\cot 2\theta \sinh \lambda & \cosh \lambda%
\end{array}%
\right) ,\;\;K=\frac{\sinh \lambda }{S}\left( 
\begin{array}{cc}
\sinh \lambda \cos 2\theta & \cosh \lambda \sin 2\theta \\ 
\cosh \lambda \sin 2\theta & -\sinh \lambda \cos 2\theta%
\end{array}%
\right) =\tilde{K},  \label{40}
\end{equation}%
and%
\begin{equation}
M^{-1}K=\left( 
\begin{array}{cc}
0 & \sinh \lambda /\sin 2\theta \\ 
\sinh \lambda /\sin 2\theta & 0%
\end{array}%
\right) ,\text{ \ }  \label{41}
\end{equation}%
\begin{equation}
M^{-1}\tilde{M}^{-1}=\left( 
\begin{array}{cc}
1+\left( \sinh \lambda /\sin 2\theta \right) ^{2} & 0 \\ 
0 & 1+\left( \sinh \lambda /\sin 2\theta \right) ^{2}%
\end{array}%
\right) .  \label{42}
\end{equation}%
One can check the unitarity of $U$ via the following commutative relations, 
\begin{equation}
\begin{array}{c}
\left[ Ua_{i}U^{-1},Ua_{j}U^{-1}\right] =\left( M^{-1}K\tilde{M}%
^{-1}-M^{-1}\left( M^{-1}K\right) ^{T}\right) _{ij}=0, \\ 
\left[ Ua_{i}U^{-1},Ua_{j}^{\dagger }U^{-1}\right] =\left[ M^{-1}\tilde{M}%
^{-1}-\left( M^{-1}K\right) \left( M^{-1}K\right) ^{T}\right] _{ij}=\delta
_{ij}.%
\end{array}
\label{43}
\end{equation}%
From (23) and (30) we know that $U$ is a new squeezing operator which
squeezes $\left\vert \eta ,\theta \right\rangle $ in a natural way, 
\begin{equation}
U\left\vert \eta ,\theta \right\rangle =\frac{1}{\mu }\left\vert \eta /\mu
,\theta \right\rangle .  \label{44}
\end{equation}

\section{The property of the squeezed state generated by U}

Writing Eq. (39) explicitly, we have 
\begin{eqnarray}
Ua_{1}U^{-1} &=&a_{1}\cosh \lambda -a_{2}\cot 2\theta \sinh \lambda
-a_{2}^{\dagger }\csc 2\theta \sinh \lambda ,  \label{45} \\
Ua_{2}U^{-1} &=&a_{2}\cosh \lambda +a_{1}\cot 2\theta \sinh \lambda
-a_{1}^{\dagger }\csc 2\theta \sinh \lambda .  \nonumber
\end{eqnarray}%
It then follows 
\begin{equation}
UX_{1}U^{-1}=\frac{1}{\sqrt{2}}U\left( a_{1}+a_{1}^{\dagger }\right)
U^{-1}=X_{1}\cosh \lambda -X_{2}\cot \theta \sinh \lambda ,  \label{46}
\end{equation}%
\begin{equation}
UX_{2}U^{-1}=X_{2}\cosh \lambda -X_{1}\tan \theta \sinh \lambda ,  \label{47}
\end{equation}%
\begin{equation}
UP_{1}U^{-1}=\frac{1}{\sqrt{2}i}U\left( a_{1}-a_{1}^{\dagger }\right)
U^{-1}=P_{1}\cosh \lambda +P_{2}\tan \theta \sinh \lambda ,  \label{48}
\end{equation}%
\begin{equation}
UP_{2}U^{-1}=P_{2}\cosh \lambda +P_{1}\cot \theta \sinh \lambda ,  \label{49}
\end{equation}%
so under the $U$ transformation the two quadratures for two-mode optical
field become 
\begin{equation}
U\left( X_{1}+X_{2}\right) U^{-1}=X_{1}\left( \cosh \lambda -\tan \theta
\sinh \lambda \right) +X_{2}\left( \cosh \lambda -\cot \theta \sinh \lambda
\right) ,  \label{50}
\end{equation}%
\begin{equation}
U\left( P_{1}+P_{2}\right) U^{-1}=P_{1}\left( \cosh \lambda +\cot \theta
\sinh \lambda \right) +P_{2}\left( \cosh \lambda +\tan \theta \sinh \lambda
\right) .  \label{51}
\end{equation}%
Using (31) we know that $U^{-1}=U^{\dagger }$ generates the $\theta -$%
related squeezed vacuum state, 
\begin{equation}
U^{-1}\left\vert 00\right\rangle =\frac{\sin 2\theta }{\sqrt{S}}\exp \{\frac{%
\cos 2\theta }{2S}\sinh ^{2}\lambda (a_{1}^{\dagger 2}-a_{2}^{\dagger 2})-%
\frac{\sin 2\theta }{2S}a_{1}^{\dagger }a_{2}^{\dagger }\sinh 2\lambda
\}\equiv \left\vert \left. {}\right. \right\rangle _{\lambda ,\theta }.
\label{52}
\end{equation}%
The expectation value of the two quadratures in the state $\left\vert \left.
{}\right. \right\rangle _{\lambda ,\theta }$ are%
\begin{equation}
_{\lambda ,\theta }\left\langle \left. {}\right. \right\vert \left(
X_{1}+X_{2}\right) \left\vert \left. {}\right. \right\rangle _{\lambda
,\theta }=0,\text{ \ }_{\lambda ,\theta }\left\langle \left. {}\right.
\right\vert \left( P_{1}+P_{2}\right) \left\vert \left. {}\right.
\right\rangle _{\lambda ,\theta }=0,  \label{53}
\end{equation}%
thus the variance of the two quadratures are 
\begin{eqnarray}
_{\lambda ,\theta }\left\langle \Delta \left( X_{1}+X_{2}\right)
^{2}\right\rangle _{\lambda ,\theta } &=&_{\lambda ,\theta }\left\langle
\left. {}\right. \right\vert \left( X_{1}+X_{2}\right) ^{2}\left\vert \left.
{}\right. \right\rangle _{\lambda ,\theta }=\left\langle 00\right\vert
U\left( X_{1}+X_{2}\right) ^{2}U^{-1}\left\vert 00\right\rangle  \label{54}
\\
&=&\cosh ^{2}\lambda +\frac{\sinh ^{2}\lambda }{2}\left( \tan ^{2}\theta
+\cot ^{2}\theta \right) -\frac{\sinh 2\lambda }{2}\left( \tan \theta +\cot
\theta \right) ,  \nonumber
\end{eqnarray}%
\begin{eqnarray}
_{\lambda ,\theta }\left\langle \Delta \left( P_{1}+P_{2}\right)
^{2}\right\rangle _{\lambda ,\theta } &=&_{\lambda ,\theta }\left\langle
\left. {}\right. \right\vert \left( P_{1}+P_{2}\right) ^{2}\left\vert \left.
{}\right. \right\rangle _{\lambda ,\theta }=\left\langle 00\right\vert
U\left( P_{1}+P_{2}\right) ^{2}U^{-1}\left\vert 00\right\rangle  \label{55}
\\
&=&\cosh ^{2}\lambda +\frac{\sinh ^{2}\lambda }{2}\left( \tan ^{2}\theta
+\cot ^{2}\theta \right) +\frac{\sinh 2\lambda }{2}\left( \tan \theta +\cot
\theta \right)  \nonumber
\end{eqnarray}%
Especially, when $\theta =\pi /4,$ this $\theta -$related squeezed vacuum
state reduces to the usual two-mode squeezed state, (54) and (55)
respectively become 
\begin{equation}
_{\lambda ,\pi /4}\left\langle \Delta \left( X_{1}+X_{2}\right)
^{2}\right\rangle _{\lambda ,\pi /4}=e^{-2\lambda },\text{\ \ }_{\lambda
,\pi /4}\left\langle \left. {}\right. \right\vert \left( P_{1}+P_{2}\right)
^{2}\left\vert \left. {}\right. \right\rangle _{\lambda ,\pi /4}=e^{2\lambda
},  \label{56}
\end{equation}%
as expected. On the other hand, due to $\tan ^{2}\theta +\cot ^{2}\theta
\geqslant 2,$ $\tan \theta +\cot \theta \geqslant 2,$ from (55) we see%
\begin{equation}
_{\lambda ,\theta }\left\langle \Delta \left( P_{1}+P_{2}\right)
^{2}\right\rangle _{\lambda ,\theta }\geqslant \left( \cosh \lambda +\sinh
\lambda \right) ^{2}=e^{2\lambda },  \label{57}
\end{equation}%
which means that the $\theta -$related squeezed vacuum state can exhibit
more stronger squeezing in one quadrature than that of the usual two-mode
squeezed vacuum state. Finally, since $\sin 2\theta \leq 1,$ $\cos
^{2}2\theta \leq 1,$ when the squeezing parameter $\mu =e^{\lambda }$ is
large enough such that $\cosh ^{2}\lambda \gg \cos ^{2}2\theta ,$ $S=\cosh
^{2}\lambda -\cos ^{2}2\theta \sim \cosh ^{2}\lambda ,$ then $U\left\vert
00\right\rangle $ is approximately equal to (up to a constant factor) 
\begin{equation}
U\left\vert 00\right\rangle \rightarrow \exp \{\frac{\tanh ^{2}\lambda }{2}%
\cos 2\theta (a_{1}^{\dagger 2}-a_{2}^{\dagger 2})+a_{1}^{\dagger
}a_{2}^{\dagger }\tanh \lambda \sin 2\theta \}\left\vert 00\right\rangle
\label{58}
\end{equation}%
Experimentally, this state can be approximately produced when two light
fields respectively squeezed in $X_{i}\ $and $P_{i}$ with the same squeezing
parameter $\mu =e^{\lambda },$ expressed by $e^{\frac{1}{2}a_{1}^{\dagger
2}\tanh \lambda }\left\vert 0\right\rangle _{1}$ and$\ e^{-\frac{1}{2}%
a_{2}^{\dagger 2}\tanh \lambda }\left\vert 0\right\rangle _{1}$
respectively, entering the asymmetric beamsplitter's two input ports and get
superimposed, then using (19) we know that the output state is 
\begin{eqnarray}
&&\exp \left[ -2i\theta J_{y}\right] e^{\frac{1}{2}a_{1}^{\dagger 2}\tanh
\lambda }e^{-\frac{1}{2}a_{2}^{\dagger 2}\tanh \lambda }\left[ 2i\theta J_{y}%
\right] \exp \left[ -2i\theta J_{y}\right] \left\vert 00\right\rangle
\label{59} \\
&=&\exp \{\frac{\tanh \lambda }{2}\left[ \left( a_{1}^{\dagger }\cos \theta
+a_{2}^{\dagger }\sin \theta \right) ^{2}-\left( a_{2}^{\dagger }\cos \theta
-a_{1}^{\dagger }\sin \theta \right) ^{2}\right] \}\left\vert 00\right\rangle
\nonumber \\
&=&\exp \{\frac{\tanh \lambda }{2}\cos 2\theta (a_{1}^{\dagger
2}-a_{2}^{\dagger 2})+a_{1}^{\dagger }a_{2}^{\dagger }\tanh \lambda \sin
2\theta \}\left\vert 00\right\rangle ,  \nonumber
\end{eqnarray}%
which is approximately equal to (58) when $\tanh ^{2}\lambda \sim \tanh
\lambda $.

In summary, as a non-trivial generalization of the fact that a 50/50
beamsplitter can produce an EPR entangled state, we see that a new entangled
state $\left\vert \eta ,\theta \right\rangle $ can be generated at the
output of an asymmetric beam-splitter with two squeezed states as inputs.
The two input states are squeezed in orthogonal quadratures while the degree
of single-mode squeezing is assumed equal for both input modes. Such states
are potentially useful, because they make up a complete and orthonormal
representation in two-mode Fock space as Dirac's theory stated. $^{17}$
Using $\left\vert \eta ,\theta \right\rangle $ we have derived new squeezed
state (52) and analysed its properties. The foundation of $\left\vert \eta
,\theta \right\rangle $ generalizes the EPR entangled state representation
with continuous variables. For the 3-mode squeezed state which relates to
the corresponding entangled state representation we refer to [18].

\subsection{Acknowledgment}

One of the authors, Hong-yi Fan, considers that this work is in memory of
Prof. L. Mandel, one of the pioneers of quantum optics, who friendly invited
him to visit University of Rochester in 1987 and discussed with him on
squeezed states, entangled states and the existence of creation operator's
eigenket.

\end{document}